\documentclass[%
reprint,
 amsmath,amssymb,
 aps,
]{revtex4-1}

\usepackage{graphicx}
\usepackage{dcolumn}
\usepackage{bm}

\usepackage{subfigure}
\usepackage{color}
\usepackage{textcomp}
\newcommand{\com}[1]{#1}
\definecolor{grey}{rgb}{0.3,0.6,0.3} 

\newcommand{\donotshow}[1]{}

\begin{document}


\title{Damage-cluster distributions and size effect on strength in compressive failure}

\author{Lucas Girard}
\email{lucas.girard@geo.uzh.ch}
\affiliation{Dep. of Geography, University of Z\"urich, Switzerland}

\author{J\'er\^ome Weiss}%
\affiliation{Laboratoire de Glaciologie et G\'eophysique de l\textquotesingle Environnement, CNRS - Universit\'e J. Fourier, Grenoble, France}

\author{David Amitrano}%
\affiliation{Institut des Sciences de la Terre, CNRS - Universit\'e J. Fourier, Grenoble, France}

\date{\today}

\begin{abstract}
We investigate compressive failure of heterogeneous materials on the basis of a continuous progressive damage model. The model explicitely accounts for tensile and shear local damage and reproduces the main features of compressive failure of brittle materials like rocks or ice. We show that the size distribution of damage-clusters, as well as the evolution of an order parameter, the size of the largest damage-cluster, argue for a critical interpretation of fracture. The compressive failure strength follows a normal distribution with a very small size effect on the mean strength, in good agreement with experiments.
\end{abstract}

\maketitle


Understanding how materials break is a fundamental problem that has both theoretical and practical relevance. The topic has received considerable renewed attention during the last few decades because of the limitations of the classical Griffith theory for heterogeneous media \citep{herrmann1990,alava2006,alava2009}. \com{The practical applications are numerous, from material and structural design, to the important problem of size effects on strength (e.g. \citep{bazant1998})}. The two key components that make material failure challenging to understand are long-range interactions and material disorder.

Traditionally, Weibull and Gumbel distributions associated with the weakest-link approach have been widely used to describe the strength of brittle materials. These distributions naturally arise from extreme-value statistics of initial defect distributions based on the assumptions that \citep{peterlik2001} (i) defects do not interact with one another, (ii) failure of the whole system is dictated by the activation of the largest flaw (the weakest-link hypothesis), and finally (iii) the material strength can be related to the critical defect size. These assumptions are reasonable for materials with relatively weak disorder loaded under tension, but do not hold for heterogeneous materials with broad distribution of initial disorder or for loading conditions stabilizing crack propagation, such as compression. In these cases there is experimental evidence that a considerable amount of damage can be sustained before failure \citep{lockner1991}. Under these conditions, failure is the culmination of a complex process involving the nucleation, propagation, interaction and coalescence of many microcracks \citep{reches1994}. \com{Stress states observed under various natural conditions, ranging from soil and rock mechanics to earthquake physics, suggest the importance of compressive failure.}


A cornerstone for the understanding of breakdown of disordered media has been lattice models of fracture in which networks with prescribed bond failure thresholds are subject to increasing external loads \citep{alava2006}. In these models, failure is described on a qualitative level as the interplay between disorder and elasticity. When strong disorder is considered, these models suggest that fracture strength does indeed not follow a Weibull or a Gumbel distribution but a log-normal distribution \citep{nukala2004strength, *zapperi2006, nukala2005}. Similar strength distributions have been obtained for different model types (fuses and springs), in 2D and 3D, suggesting that it is a general feature of failure in heterogeneous materials with broad disorder \citep{alava2006}.

On a more fundamental point of view, the evolution of the distribution of crack-cluster sizes in the vicinity of failure has implications on the interpretation of rupture as a phase transition. For a second-order or critical phase transition, local quantities such as the size of crack-clusters are expected to show scaling and a diverging correlation length, such as in the percolation problem \citep{herrmann1990}. In the limit of infinitely strong disorder, fracture can be mapped onto the percolation problem \citep{roux1988}, suggesting a critical transition. This interpretation remains however controversial in the case of non-infinite disorder, as lattice models show an abrupt localization of damage at failure, without a diverging correlation length, arguing instead for a first-order transition \citep{nukala2005, zapperi1999,nukala2004}.

In this letter, we revisit these problems for compressive failure of an heterogeneous material with variable range of disorder, on the basis of a continuous progressive damage model \citep{amitrano1999, girard2010}. This model is more realistic than lattice models \com{or scalar damage models \citep{zapperi1997}}, as it explicitely accounts for the tensorial nature of stresses and strains. The principal features of compressive failure of brittle materials like \com{low-porosity} rocks or ice are captured by the model: the macroscopic stress-strain response or the progressive localization of damage onto an inclined fault at failure \citep{lockner1991, iliescu2004, *Kat-Rec-04}. \com{A detailed comparison of the model with experimental results of compressive failure of rock samples has been performed using acoustic emission and damage avalanche statistics \citep{amitrano2003}}. Recently, an analysis of this damage localization, either tracked from damage avalanches or from the evolution of continuous strain-rate fields, showed a divergence of the associated correlation length towards the failure, i.e. argued for the critical point hypothesis \citep{girard2010}. Here we report that the size distribution of damage clusters as well as the evolution of an order parameter, the size of the largest damage cluster, also argue for a critical interpretation of compressive fracture with specific scaling laws in the pre- as well as post-failure phases. We also show that the compressive failure strength has a normal distribution and characterize the associated size effect.


The model, described in more details elsewhere \citep{amitrano1999, girard2010}, considers a continuous 2D elastic material (Hooke's law) under plane stress, with progressive local damage. Damage is represented by a reduction of the isotropic elastic modulus $Y_i$ of the element $i$, $Y_i(n+1) = Y_i(n) d_0$ with $d_0=0.9$, each time the stress state on that element exceeds a given threshold. This elastic softening simulates an increase in crack density at the element scale \citep{kachanov1994} \com{as supported by experiments \citep{cox1993}. In high porosity materials, modifications of elastic properties can also result from local compaction (e.g. \citep{katsman2005}), a problem not considered here.} The stress field is recalculated each time a damage event occurs by solving the equation of static equilibrium using a finite element scheme. As the result of elastic interactions, the stress redistribution following a damage event can set off an avalanche of damage, which stops when the damage threshold is no longer fulfilled by any element.


The Coulomb criterion, $\tau = \mu \sigma_N+C$, of wide applicability for brittle materials under compressive stress states \citep{weiss2009coulomb, *jaeger1979}, defines the damage threshold. $\tau$  and $\sigma_N$ are respectively the shear and normal stress on the element (sign convention positive in compression), $\mu$ is an internal friction coefficient identical for all elements, whereas quenched disorder is introduced through the cohesion $C$ randomly drawn from a uniform distribution. We use $\mu=0.7$, a common value for most geomaterials \citep{weiss2009coulomb, *jaeger1979}. This envelope is completed by a truncation in tension in the Mohr's plane, i.e. the element is damaged if $\sigma_N=\sigma_{Ntensile} =-2\times 10^{-3}\times Y_0$. The simulations, which start with undamaged material ($Y_i=Y_0=\mbox{const.}$), are performed on rectangular meshes of randomly oriented triangular elements. A uniaxial compression loading is applied by increasing the vertical displacement of the upper boundary of the system  (i.e. strain-driven loading), whereas the lower boundary is fixed and left and right boundaries can deform freely. The loading increment is extrapolated to damage the weakest element, ensuring infinitely slow driving compared to stress redistribution time. Two series of simulations, with meshes of linear size $L$ varying from 8 to 128 elements, were performed with different ranges of disorder: $0.5\times 10^{-3}Y_0 \leq C \leq 10^{-3}Y_0$ which we refer to as the H1 disorder, $0.2\times 10^{-3}Y_0 \leq C \leq 10^{-3}Y_0$ refered to as H2. The range of disorder used in this study is thus narrower than in lattice models where the failure threshold distribution usually extends down to $0$ (e.g. \citep{nukala2004strength}). The number of independent simulations performed with each system size is $10^4$ up to $L=16$, $5\times10^3$ for $L=32$, $10^3$ for $L=64$ and $10^2$ for $L=128$.

\begin{figure} \center
\includegraphics[width=0.5\textwidth]{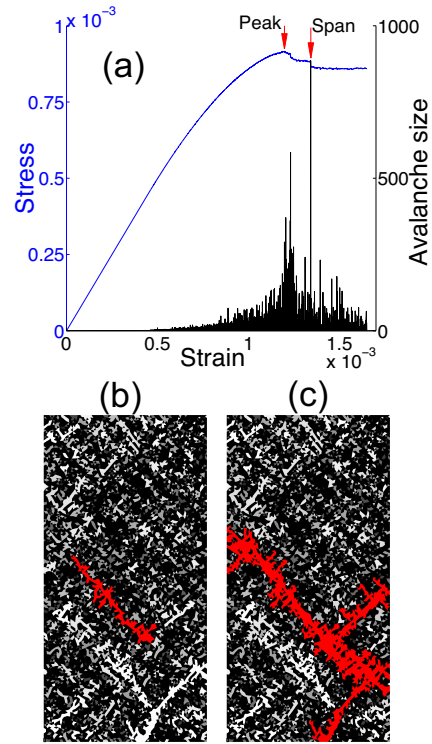}
\caption{\label{fig:macro}(a) Example of the evolution of the macroscopic stress versus strain, as well as the damage avalanche size, i.e. number of damage events per iteration ($L=64$, H2 heterogeneity). Red arrows indicate respectively (i) peak load and (ii) the point where the largest damage cluster spans the system. Maps of damage clusters at these two points are respectively plotted as (b) and (c) with the largest cluster in red.}
\end{figure}

In the early stages of deformation, damage is scattered homogeneously and the macroscopic stress-strain response remains essentially linear (Fig. \ref{fig:macro}(a)). As deformation proceeds, macroscopic softening occurs up to a maximum stress, the strength or peak load, followed by one or a few macroscopic stress drops. Then, the macroscopic stress remains approximately constant, i.e. the behaviour mimics plasticity.

First, we identify damage clusters whose size $A$ is defined as the total surface area of adjacent elements (i.e. sharing two nodes) that have been damaged at least one time from the beginning of the simulation. At peak load, the largest damage cluster does not yet span the system (Fig. \ref{fig:macro}(b)). Then, during the post-peak phase, damage events are localized in the vincinity of one or a few large damage clusters which eventually evolve into a spanning cluster, connecting two opposite boundaries of the system (Fig. \ref{fig:macro}(c)). Using similar simulations, we previously reported \citep{girard2010} that the spatial correlation length associated with damage events or strain localization reaches the size of the system at peak load. This means that the divergence of the dynamical correlation length associated with damage avalanches and strain localization precedes the geometrical evolution of the fault. In other words, due to the long-range correlations in the stress field that emerge in the vicinity of failure, damage clusters do not need to be interconnected into a spanning cluster for the global failure to occur. Strain-driven compressive failure experiments of rocks have reported similar observations: the failure plane is not fully formed at peak load and only appears in the post-peak phase with acoustic emissions strongly localized along the main fault during this phase \citep{lockner1991}. This agreement with experiments indicates that the model is relevant to describe damage evolution in the post-peak phase.

In what follows the distance to the peak load, identified as the critical point, is tracked in terms of a control parameter $\Delta = \left| \epsilon_{mp} -\epsilon_m \right| / \epsilon_{mp}$, where $\epsilon_m$ is the macroscopic strain, reaching $\epsilon_{mp}$ at peak load. This definition yields $\Delta = 1$ at the first damage event, decreasing down to $\Delta = 0$ at the peak load and increasing again after peak load.

\begin{figure} \center
\includegraphics[width=0.5\textwidth]{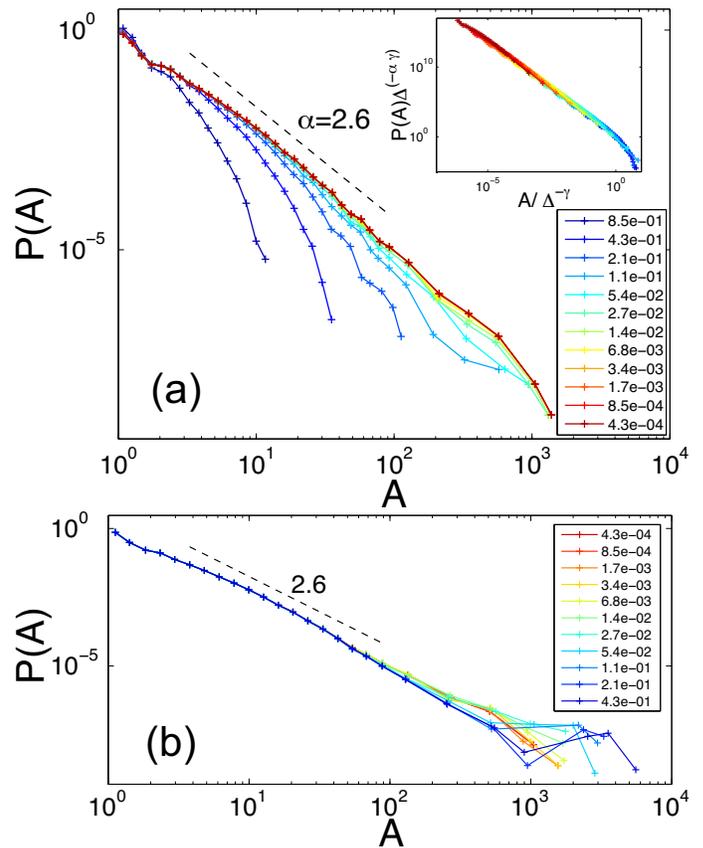}
\caption{\label{fig:pdfcluster}(a) Probability density function (PDF) of the size of damage clusters $A$ for bins of the control parameter $\Delta$ before peak load, and associated data collapse (inset). (b) PDF of $A$ after the peak load ($L=128$, H2 disorder). The color code corresponds to the value of $\Delta$ as given by the legend, red is closest to peak load.}
\end{figure}

We analyse the distribution of damage cluster sizes $A$: before peak load it follows a power law with an exponential cut-off at large sizes that increases in the vicinity of peak load (Fig. \ref{fig:pdfcluster}), $P(A) \sim A^{-\alpha} \exp(-A/{A}^*)$. For finite-size systems, the evolution of ${A}^*$ is controled by the system-size itself (finite size effect) as well as the distance to peak load. We make the simplifying hypothesis that, for the largest system ($L=128$), the cut-off size is only a function of the control parameter, ${A}^* \sim \Delta^{-\gamma}$. This allows us to estimate the value of $\alpha$ and $\gamma$ through a data collapse (Fig. \ref{fig:pdfcluster}, inset). For the H1 disorder we find $\alpha=3.6\pm0.1$ and $\gamma=1.3\pm0.2$, while for H2, $\alpha=2.6\pm0.1$ and $\gamma=1.6\pm0.2$. 

The nature of the distribution identified argues for a critical interpretation of failure, since we show here that a local quantity, the damage cluster size, shows scaling in the vicinity of failure. Moreover, this evolution is related to the divergence of the correlation length and the critical exponents can be related geometrically: since $A$ is the surface area of damage clusters, it can be expected to scale with the correlation length $\xi$ and the fractal dimension of damage clusters $D$ as ${A}^* \sim \xi^D$. In Ref. \citep{girard2010} we reported from an analysis of strain-rate fields $\xi \sim \Delta^\lambda$, with $\lambda=1.0\pm 0.1$, $D=1.15\pm0.05$ for H1, and $D=1.4\pm0.1$ for H2. The geometrical relationship $A^* \sim \Delta^{\lambda D}$ yields $\gamma=\lambda D$ for large system sizes. When accounting for the uncertainty on the exponent values, the later relationship is captured for the largest system size ($L=128$), with a correlation length exponent independent of disorder. The analysis can also be refined to account for the finite size effect on $A$ \citep{suppmat}. These results contrast with the log-normal distribution of crack-cluster sizes observed for lattice models of fracture \citep{nukala2006crack}.

\begin{figure} \center
\includegraphics[width=0.5\textwidth]{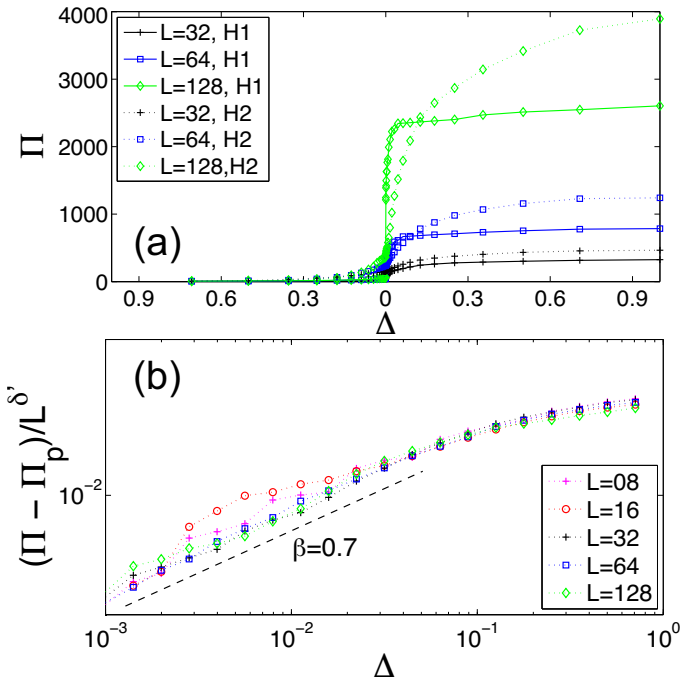}
\caption{\label{fig:PI} (a) Growth of the largest damage cluster size $\Pi$, $\Delta=0$ corresponds to peak load, left and right of this point are respectively the pre- and post-peak phases. (b) Data collapse of the largest cluster size $\Pi$ after the peak load (H2 disorder).}
\end{figure}

After peak load, the distribution of cluster sizes shows a power law with similar exponent, as well as an additional peak at large sizes that progressively grows as the distance to failure increases (Fig. \ref{fig:pdfcluster}). This suggests that the physics governing the evolution of damage is different during this phase and favors the growth of the largests clusters. As the changes in the distribution are concentrated in its tail during this phase, we analyze the growth of the largest damage cluster, which size is denoted $\Pi$. By analogy with the percolation theory, we consider $\Pi$ as the order parameter of the phase transition, defined here for the entire deformation history. The general evolution of $\Pi$ is continuous with an inflexion point at failure (Fig. \ref{fig:PI} (a)). The degree of initial heterogeneity shows a large influence on the behavior around peak load, a narrower heterogeneity leading to a more abrupt transition. At peak load, $\Pi$ scales with the system size $\Pi(L, \Delta \to 0) = \Pi_p(L) = L^\delta$, with  $\delta=0.3\pm0.1$ for H1 and $\delta=0.8\pm0.1$ for H2. This means that with a broader disorder the largest cluster is in average closer to span the system at peak load.

Next, we analyze the growth of the largest cluster in the post-peak phase. The net growth of $\Pi$ scales with the system size as well as with the control parameter following: 
\begin{equation}
(\Pi - \Pi_p) \sim L^{\delta'} \Delta^{\beta}.
\label{eq:piscaling}
\end{equation}
The value of $\delta'$ is estimated from a data collapse, yielding $\delta'=1.3\pm0.1$ (H1) and $\delta'=1.9\pm0.1$ (H2) (Fig. \ref{fig:PI} (b)). In the vicinity of peak load, the resulting collapsed curve shows a power law yielding $\beta=0.4$ (H1) and $\beta=0.7$ (H2).

\begin{figure} \center
\includegraphics[width=0.5\textwidth]{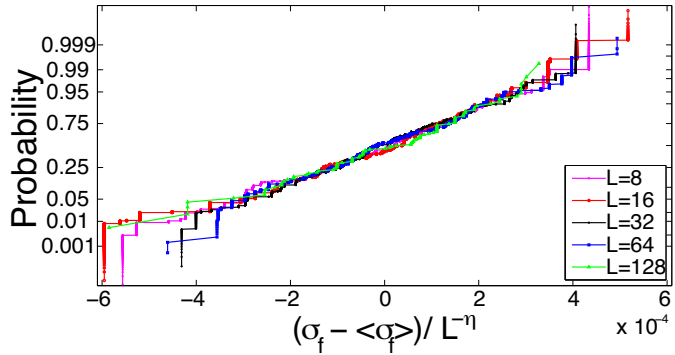}
\caption{\label{fig:strength}Normal probability plot of the strength distribution. Albeit minute deviation, the plot shows a collapsed straight line, as expected for a normal distribution ($L=128$, H2). The figure also demonstrates the power law scaling of the standard deviation $std(\sigma_f) \sim L^{-\eta}$.}
\end{figure}

We established that the growth of clusters during this post-peak phase is equally contributed by the coalescence of existing clusters as well as by the extension of the total damaged area, since the two mechanisms are described by a scaling law equivalent to Eq. \ref{eq:piscaling}, with similar exponents \citep{suppmat}. The two mechanisms are tightly linked during this phase, as the expansion/branching of clusters bridges gaps with other clusters and induces coalescence. Close to peak load, we observed that Eq. \ref{eq:piscaling} also applies to the growth of other large damage clusters, i.e. the 2nd, 3rd, 4th largests clusters for example, until they eventually coalesce or stop growing as distance to peak load increases. 

On Fig. \ref{fig:PI} (b), the deviation from the power law for $\Delta>0.1$ corresponds to the point where the largest cluster becomes spanning. Beyond this point the geometrical growth of the largest cluster appears progressively inhibited and the power law scaling does not hold, i.e. further deformation is accomodated only by additional damage and shear along the inclined, mature fault, in full agreement with observations \citep{lockner1991}.


Finally, we analyse the compressive failure strength, which we define as the maximal macroscopic stress $\sigma_{f}$ reached during a simulation. We verified that the strength distribution cannot be represented by a Weibull nor by a Gumbel distribution \citep{suppmat}, meaning that the weakest-link approach is irrelevant to compressive failure. Instead, the strength distribution can be described by a normal distribution (Fig. \ref{fig:strength}). Fig. \ref{fig:strength} also shows that the standard deviation of the strength scales as $std(\sigma_f) \sim L^{-\eta}$, where a data collapse is obtained for $\eta=0.4$ for H1 disorder and $\eta=0.65$ for H2. The mean strength shows a very slow decrease with increasing system size that can be represented as $\left< \sigma_f \right> \sim L^{-\theta}$ where $\theta=0.02$. This is in excellent agreement with compressive failure experiments on rock and ice that show no significant sample size effect on the mean strength \citep{kuehn1993, *tsur1982}.


To conclude, we showed that in compressive failure, the peak load can be considered as a critical point regarding the evolution of damage clusters. Scaling laws, specific to the pre- and post-peak phases, were shown to describe the evolution of the size distribution of damage clusters. This expresses the difference in the physics that govern the growth of damage clusters in these two phases. Compressive failure thus appears as a complex cumulative process involving long-range correlations, interactions and coalescence of microcracks. In such conditions, the hypotheses of the weakest-link approach, describing failure as an abrupt fist-order transition, are violated and the strength distributions predicted by extreme-value statistics do not apply at all.
In lattice models of fracture, strength distributions are neither captured by Weibull nor by Gumbel distributions \citep{nukala2004strength, *zapperi2006, nukala2005}. An interesting point is that the ranges of initial disorder that we have considered are not as broadly distributed as in lattice models. This suggests that for loading modes stabilizing damage propagation, such as compression, even under narrowly distributed initial disorder, the weakest-link approach does not hold and failure can be interpreted as a critical phase transition.

All computations were performed at SCCI-CIMENT Grenoble.


\bibliography{LM13290_revision}


\newpage
\clearpage

\end{document}